\documentclass[prl,twocolumn,showpacs]{revtex4}
\usepackage{graphicx,bm,amssymb,amsmath}

\newcommand{\pdag}{{\phantom{\dagger}}}

\newcommand{\beq}{\begin{equation}}
\newcommand{\eeq}{\end{equation}}

\newcommand{\beqa}{\begin{eqnarray}}
\newcommand{\eeqa}{\end{eqnarray}}

\newcommand{\PRB}[1]{Phys. Rev. B~\textbf{#1}}

\newcommand{\RMP}[1]{Rev. Mod. Phys.~\textbf{#1}}

\begin{document}

\title{Kondo screening cloud in a one dimensional wire: Numerical
  renormalization group study}

\author{L\'aszl\'o Borda}
\affiliation{Department of Theoretical Physics
and 
Research Group ``Theory of Condensed Matter'' of the Hungarian Academy of
Sciences,\\ 
Budapest University of Technology and Economics,
Budafoki \'ut 8.  H-1521 Budapest, Hungary}

\date{\today}

\begin{abstract}
We study the Kondo model --a magnetic impurity coupled to a one dimensional 
wire via exchange coupling-- by using Wilson's numerical renormalization group
(NRG) technique. By applying an approach similar to which was used to compute 
the two impurity problem we managed to improve the bad spatial resolution of
the numerical renormalization group method. 
In this way we have calculated the impurity spin - conduction
electron spin correlation function which is a measure of the Kondo
compensation cloud whose existence has been a long standing problem
in solid state physics.
We also present results on the temperature dependence
of the Kondo correlations.
\end{abstract}

\maketitle

{\em Introduction}---
As being one of the most interesting quantum impurity problems, the
Kondo effect\cite{kondo} --when a localized impurity spin 
interacts with 
the
itinerant electrons via spin exchange coupling-- has been studied
for decades both theoretically and experimentally\cite{hewson} 
and has recently 
come to its renaissance when it has been predicted to appear\cite{qd_th} 
and
observed\cite{qd_ex} 
in quantum dot systems. Even though the Kondo effect seems
to be well-understood now, still, there is a controversial question
which has not been clarified yet: Whether there exists 
%a large 
an extended
``Kondo screening cloud'' 
%characterized by the length scale
%$\xi_K=\hbar v_F/k_BT_K$ ($v_F$ is the Fermi velocity $T_K$ is the
%Kondo temperature)
or not?

In case of the single channel Kondo problem --when the local spin 
%interacts with 
is coupled to
one band of conduction electrons-- the 
%coupling to
interaction with
the electrons lifts the ground state degeneracy of the isolated
spin resulting in an effective ``screening'' of the impurity spin
below a certain energy scale, the so-called Kondo temperature
$T_K$. It is adequate to ask the question what is the typical 
length scale in which the impurity spin gets screened. 
As a simple estimate, just by comparing 
the scales of the competing kinetic and binding energies, one 
easily gets a length scale, the so-called Kondo coherence length,
$\xi_K=\hbar v_F/k_BT_K$. By substituting the typical Fermi velocity
$v_F$ in metals, and 
%assuming 
taking a typical Kondo temperature
$T_K\sim 1K$, one gets
$\xi_K\sim 1\mu m$, which is comparable with, or even larger than
the typical dimension of today's mesoscopic devices. However, to 
measure that screening cloud is highly non-trivial from
experimental point of view.

In order to observe the screening cloud, one has to measure 
{\em correlations} between the impurity spin and the conduction
electron spin density. It is very difficult to imagine such kind
of a measurement in bulk metallic samples. Recently, it has become
possible to perform scanning tunneling microscopy 
(STM) measurements on
metallic surfaces addressing a single magnetic impurity and its 
close neighborhood\cite{STM}. 
In those experiments the screening cloud was
not observed since the STM tip probes the local {\em charge} density of
states which is affected by the Kondo correlations
in the close neighborhood of the impurity only. Even in
case of having spin-polarized STM tips, it is very difficult to
measure the spin {\em correlations} since the Kondo effect 
manifests itself in continuous, coherent spin flip processes at
a time scale $\sim 1/T_K$ meaning that one has to measure at a 
frequency of tens of gigahertzs. The idea of suppressing the
impurity spin fluctuations by applying a local magnetic field is
doomed to fail since a magnetic field which is large enough to
suppress the spin fluctuations is necessarily large enough to
destroy Kondo correlations as well.

There were experimental setups proposed in order to measure
the Kondo cloud in {\em confined systems}\cite{Balseiro}, 
e.g. in case of a quantum
dot attached to {\em one dimensional}
lead with length shorter than, or comparable with 
$\xi_K$. Such kind of a proposal suffers from the facts that
it is destructive, i.e. the Kondo effect disappears
for 
{\em one dimensional}
leads 
shorter than $\xi_K$ and one can still argue that
the suppression of the Kondo effect is a result of having the
level spacing in the lead $\delta\epsilon>T_K$ for short enough 
leads therefore one does not need the theoretical construction of
the screening cloud to explain the experimental findings.
In case of an impurity embedded into a higher dimensional
environment, another length scale $l_K$ emerges\cite{box}
by equating $\Delta(l_K)=T_K$ where $\Delta(l_K)$ stands for the
mean level spacing in a box with size $l_K$. For the case of
$D\geq 2$ that length scale can be substantially smaller than
$\xi_K$. In 1D systems, which are in the focus of the present paper,
the two length scales are essentially equal.

Despite those challenges, there were proposals
published to observe the Kondo screening cloud. These proposals
deal with the Knight shift\cite{knight}, persistent current\cite{persistent}
or conductance of mesoscopic systems\cite{conductance}. 

Very recently, Hand and his coworkers
established a proportionality between the weight of Kondo resonance and
the spatial extension of the Kondo correlations in mesoscopic systems
\cite{Hand}. This fact allows another, 
spectroscopic way to observe the Kondo screening cloud. 
Those proposals are very promising even though the experimental 
realizations are yet to come.

In the
present paper 
we are less ambitious since
our goal is not to propose new ways of observation but to
extend the powerful and numerically exact  
NRG\cite{Wilson}
to 
be capable to compute
spatial correlations and therefore make predictions for the outcome
of future experiments.

{\em Model and definition of correlation functions}---
The Hamiltonian of the Kondo model can be written as 
\beq
H=J\vec{S}\vec{\sigma}(0)+\sum\limits_{k,\mu}\varepsilon_kc^\dagger_{k\mu}c^\pdag_{k\mu}\;,
\label{eq:hamiltonian}
\eeq
where $\vec{S}$ is the impurity spin ($S=1/2$, located at the
origin), 
$c^\dagger_{k\mu}$ ($c^\pdag_{k\mu}$) creates (annihilates)
a conduction electron with momentum $k$ and spin $\mu$, while
$\vec{\sigma}(0)=\psi^\dagger(0)\vec{\sigma}\psi(0)$ 
stands for the electron spin density at the
position of the impurity. As it is mentioned above, the system
described by Eq.(\ref{eq:hamiltonian}) forms a singlet below the
Kondo temperature $T_K\sim D_0\exp{(-1/2\varrho_0 J)}$, where $D_0$ is the
high energy cutoff 
and $\varrho_0$ is the density of states of the
conduction electrons. 
(For sake of
simplicity, we consider a band with a constant  
density of states in the following.
Moreover, we restrict ourselves for 1D electrons because 
it is easier to treat them numerically. When it is adequate, we will
comment the case of higher dimensions as well.) A quantity which
measures the spatial extension of the singlet is the equal-time
correlator of the impurity spin and conduction electron spin density  
at position $x$: 
\begin{equation}
\chi_{t=0}(x)=\langle\vec{S}\vec{\sigma}(x)\rangle_{t=0}.
\label{eq:chi_t0}
\end{equation}
Since we know that the ground state of the Hamiltonian
Eq.(\ref{eq:hamiltonian}) is a singlet, we can easily derive a sum
rule for the equal time correlator at zero temperature:
\beq
\int\limits_{0}^{\infty}\langle\vec{S}\vec{\sigma}(x)\rangle_{t=0}
{\rm dx}=-\frac{3}{4}\;.
\label{eq:sumrule}
\eeq
 
{\em Method---}
To calculate 
spatial correlations
is a non-trivial task since
most of the methods used to investigate the Kondo model are not able
to reproduce correlation functions.
Comparatively only a very little of the theoretical study has been 
focused on spatial correlations: Perturbative calculations have
been performed\cite{pert,barzykin} as well as Monte Carlo analysis\cite{hirsch}. 
In the present paper we extend 
%Wilson's 
NRG to compute 
the Kondo correlations.

\begin{figure}
\includegraphics[width=0.9\columnwidth]{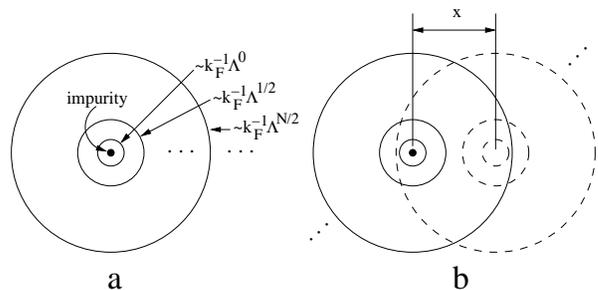}
\caption{
(a) Spherical shells in real space depicting the extents of
wave functions of states which are represented as on-site
states on the Wilson chain. As shown,  
the 
NRG
%numerical renormalization group 
method
has a good spatial resolution
around the impurity only. (b) A straightforward extension of 
%numerical renormalization group 
NRG
tackles
that problem providing a good spatial resolution at the impurity site as well
as at another, freely chosen point by introducing a second electron field. 
}
\label{fig:NRGshells}
\end{figure}

Wilson's 
%numerical renormalization
%group method 
NRG
--the most useful numerically exact method to obtain
correlation functions\cite{theo}-- 
suffers from the very bad spatial resolution
%at length scales bigger than $1/k_F$. 
away from the impurity.
This fact is a direct
consequence of the cornerstone of the method, the logarithmic 
discretization of the conduction band. In Wilson's 
%numerical renormalization group 
NRG
technique one introduces a discretization
parameter, $\Lambda$ and with the help of that discretizes the 
conduction band logarithmically by dividing it into intervals
(i.e. the $n$th interval is $]-D_0\Lambda^{-n};-D_0\Lambda^{-n-1}]$ 
for negative, and $[D_0\Lambda^{-n-1};D_0\Lambda^{-n}[$ for positive
energies measured from the Fermi energy and
$D_0$ stands for the half bandwidth)
and keeping one mode per interval only. As a next step, one maps 
the problem onto a semi-infinite chain with the impurity at the end
by means of a L\'anczos-transformation. As a 
%direct 
consequence of the
logarithmic discretization, the hopping amplitude falls off exponentially
along the chain allowing us to diagonalize the problem iteratively. As
it is shown in Ref.\cite{Wilson} the states represented as on-site states
on the Wilson-chain correspond to extended states in real space with
a typical spatial extension 
\begin{equation}
r_N\sim\frac{1}{k_F}\Lambda^{N/2}\;,
\end{equation}
where $N$ is the site index along the chain. It is obvious that 
the numerical renormalization group method
has a good spatial resolution at the position of the impurity only.
[See Fig.\ref{fig:NRGshells}a.]

%%%%%%%%%%%%%%%%%%%%%%%%%%%%%%%%
%%%%%%%%%%%%%%%%%%%%%%%%%%%%%%%%

To tackle that problem, we introduce two electron fields, one is
located at the impurity site, $\psi_0=\psi(0)$ and another one
at point $x$ where we want to evaluate the correlation function,
$\psi_1=\psi(x)$.[See Fig.\ref{fig:NRGshells}b.] 
(A very similar procedure was applied in the
numerical renormalization group
treatment of the two impurity problem\cite{2imp}.) Of course, these
fields are not orthogonal, i.e. they do not fulfill the canonical 
anticommutational relations. As a next step, we can introduce
the 
%even/odd 
proper linear combinations of these
fields, 
\begin{equation}
\psi_{\pm}=1/\sqrt{2}(\psi_0\pm\psi_1)
\end{equation}
which are now anticommuting but the 
%even/odd 
corresponding
density of states
is modified i.e. it acquires energy dependence, e.g. for 1D electrons
$\varrho_{\pm}^{1D}(\varepsilon)=\varrho_0/2[1\pm\cos(\varepsilon x/v_F)]$.
(In 2D the density of states reads 
$\varrho_{\pm}^{2D}(\varepsilon)=\varrho_0/2[1\pm J_0(\varepsilon r/v_F)]$
where $J_0$ is the zeroth Bessel function, while in 3D
$\varrho_{\pm}^{3D}(\varepsilon)=\varrho_0/2[1\pm \sin(\varepsilon
  r/v_F)/(\varepsilon r/v_F)]$. As we see, the
oscillations of the even/odd density of states decay as
$\sim r^{-(D-1)/2}$, where $D$ is the dimensionality of the problem.)
In the new basis the Kondo Hamiltonian reads
\beqa
H&=&\frac{1}{2}J\vec{S}(\psi_+^\dagger+\psi_-^\dagger)
\vec{\sigma}(\psi_++\psi_-)\nonumber\\
&+&\sum\limits_{{i=\pm}{\mu=\uparrow,\downarrow}}
\int d\varepsilon\varrho_i(\varepsilon)\varepsilon
c_{i\varepsilon\mu}^\dagger c_{i\varepsilon\mu}\;.
\eeqa
In this representation 
$\vec{\sigma}(x)=1/2[\psi_+^\dagger-\psi_-^\dagger]\vec{\sigma}[\psi_+-\psi_-]$
is represented in a high accuracy. In brief, 
the main idea behind the above transformation is that
one can get a very good
spatial resolution at the impurity site as well as at another
freely chosen position $x$ if one is willing to pay the
price of (a) having two electron channels mixed by the interaction,
(b) having energy dependent density of states and 
(c) having a separate NRG iteration for every different $x$ values.
Those difficulties are possible to be handled: The most serious one
amongst them
is the extension of NRG to arbitrary  density of states which has
been solved by Bulla et al.\cite{Ralf}
%%%%%%%%%%%%%%%%%%%%%%%%%%%%%%%%
\begin{figure}
\includegraphics[width=0.9\columnwidth]{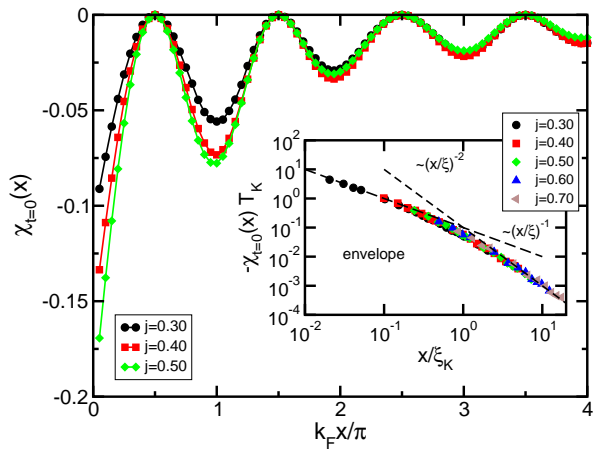}
\caption{(Color online)
The equal time spin-spin correlation function, 
$\chi_{\rm t=0}(x)=\langle\vec{S}\vec{\sigma}(x)\rangle$ as a
function of the distance $x$ measured from the impurity for
different values of the Kondo coupling $j=\varrho_0J$.
$\chi_{\rm t=0}(x)$ oscillates as $\sim\cos^2(k_Fx)$. As shown in the inset
the envelope of the oscillating part 
(i.e.$\chi_{\rm t=0}(x=n\pi/k_F)$ where $n$ is a integer)
crosses over from $\sim x^{-1}$ to
$\sim x^{-2}$ at around the Kondo coherence length, $\xi_K$. The envelope
function for different couplings nicely collapse into one universal curve.
}
\label{fig:equalt}
\end{figure} 
%%%%%%%%%%%%%%%%%%%%%%%%%%%%%%%%

{\em Results---}
In this NRG scheme the calculation of the equal time correlation
function is rather simple: $\chi_{t=0}(x)$ appears to be a static
thermodynamic
quantity which can be evaluated with a high precision. The results
for different Kondo couplings are shown in Fig.\ref{fig:equalt}:
the correlation function oscillates as $\sim\cos^2(k_Fx)$. 
As it is shown in the inset of Fig.\ref{fig:equalt},
the envelope of the correlation function for different
Kondo couplings nicely collapse into one universal curve
(apart from the points $x\leq\pi/k_F$ which show non-universal,
coupling-dependent behavior, not plotted in the inset).
The envelope
curve 
(i.e.$\chi_{\rm t=0}(x=n\pi/k_F)$ where $n$ is an integer)
changes its decay from $\sim1/x$ to $\sim1/x^2$ at 
around $\xi_K$. 
Our results supports the heuristic picture of the Kondo ground state:
The impurity is locked into a singlet with an electron whose
wave function is spread out over a distance of $\xi_K$. If we test
the correlations on a length scale much shorter than $\xi_K$ 
the impurity appears to be unscreened and a perturbative analysis can be
performed\cite{barzykin} yielding $\chi_{\rm t=0}\sim1/x$. On the
other hand, the impurity is screened
for distances $x\gg\xi_K$ and Nozi\`eres' Fermi liquid theory
applies which predicts $\sim1/x^2$ decay in 1D\cite{ishi}. 
There is, however,
a 
%relatively 
rather
wide crossover regime for which there is no reliable analytic
method of calculating 
%this correlation function. 
$\chi_{\rm t=0}(x)$.
From perturbative aspect
this is the region where the logarithmic corrections to $\sim1/x$
becomes large\cite{barzykin}. 
Alternatively, following the argumentation of the
authors of Ref.\onlinecite{Hand}, this is the region where the
impurity can be described by a fluctuating spin resulting in 
exponential decay which is overriden by the $\sim1/x^2$ decay
for even larger distances.

%%%%%%%%%%%%%%%%%%%%%%%%%%%%%%%%
\begin{figure}
\includegraphics[width=0.9\columnwidth]{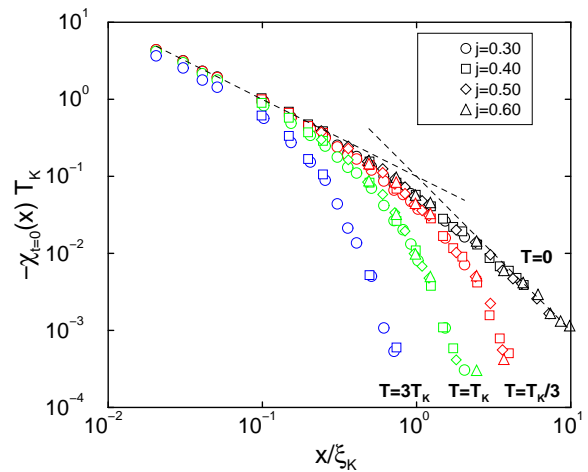}
\caption{(Color online)
The envelope of $\chi_{t=0}(x)$ for different couplings and different
temperatures: Any finite temperature introduces another energy scale,
$\xi_T=\hbar v_F/k_BT$ at which the envelope function crosses over 
from an algebraic to an exponential decay.
}
\label{fig:equalt_finiteT}
\end{figure} 
%%%%%%%%%%%%%%%%%%%%%%%%%%%%%%%%

To plot the universal curve in the inset of Fig.\ref{fig:equalt} 
we have chosen the definition of $T_K$ based on the NRG flow of the 
%lowest lying 
energy levels: We identified $T_K$ as the energy scale
at which the energy of the first excited state reaches $80\%$ of
its fixed point value. 
We have taken this definition since 
the energy spectrum is the most basic and most reliable result of any
NRG calculation and --as a  consequence of the 
universality of the Kondo effect-- the value of $T_K$ 
obtained in this way can differ from
that of other definitions up to a constant multiplying factor only.

Note that the results shown in Fig.\ref{fig:equalt}
correspond to 1D: in the case of higher dimensions the correlation
function tends to zero even faster since the sum rule
(Eq.\ref{eq:sumrule}) fixes its integral. This fact 
enters the numerical calculation through the 
$\sim r^{-(D-1)/2}$
decay of density of states oscillations and
makes the
calculation of $\chi_{t=0}(x)$ challenging for dimensions higher than $D=1$.

Up to this point we have considered the case of
zero temperature $T=0$. However, it is
known that the temperature plays an essential role in Kondo physics:
Kondo correlations are destroyed when the temperature reaches the
%scale of the 
Kondo 
%temperature 
scale
$T\sim T_K$. 
It is adequate to ask the
question how does this fact show up in $\chi_{t=0}(x)$?

Since $\chi_{t=0}(x)$ is a static quantity we obtain its complete
temperature dependence by numerical renormalization
as the iteration proceeds.
In Fig.\ref{fig:equalt_finiteT} we show the results for $T=0$,
$T=T_K/3$, $T=T_K$, and $T=3T_K$. As it is transparent from
the curves, the effect of finite temperature shows up in the 
appearance of another length scale, the thermal length scale,
$\xi_T=\hbar v_F/k_BT$. For $x<\xi_T$ the correlation function
$\chi_{t=0}(x)$ is not much affected while for $x>\xi_T$ the
correlations are cut off exponentially.

Based on those results we can now interpret the role of finite 
temperature in Kondo screening as follows: The fact that the
impurity spin is perfectly screened at $T=0$ is reflected in the sum
rule given by Eq.(\ref{eq:sumrule}). At any finite temperature the
integral is reduced and can approximately rewritten as
$\int_0^{\xi_T}\chi_{t=0}(x)dx$. When 
the temperature is lower than the Kondo temperature
$T<T_K$, the corresponding thermal
length scale is larger than $\xi_K$
Since $\chi_{t=0}(x)$ has its most weight in the region $x<\xi_K$,
the effect of a small temperature is just a small correction
to the perfect screening obtained at $T=0$. Given that in the
$x>\xi_K$ regime $\chi_{t=0}(x)\sim x^{-2}$, the correction
appears to be linear in $T$. In contrast, for $T>T_K$ the
corresponding $\xi_T$ is shorter than $\xi_K$ and the correction
to the integral is not small and consequently, we cannot speak about 
the screening of the local moment any more. 

{\em Conclusions---}
We have shown that Wilson's NRG technique is capable to
handle spatial correlations if we apply a 
straightforward
extension.
To demonstrate that we have computed the
%$\chi_{t=0}(x)=
$\langle\vec{S}\vec{\sigma}(x)\rangle_{t=0}$
correlator for a one dimensional Kondo system and 
shown
that
the decay of spin correlations crosses over from
$\sim x^{-1}$ to  $\sim x^{-2}$ at around 
the Kondo coherence length,
$\xi_K=\hbar v_F/k_BT_K$. We have shown
by calculating the temperature dependence that any finite temperature
introduces a new energy scale beyond which the Kondo correlations vanish
exponentially. Our method is --apart from the numerical challenges-- easy to
generalize to other impurity models. A very attractive example would be
the two channel Kondo model to see how the non-Fermi liquid nature of the
ground state shows up in the spatial spin correlations.

\begin{acknowledgments}
The author acknowledges the fruitful discussions with
G. Zar\'and, R. Bulla, D. Goldhaber-Gordon, A. Zawadowski
and O. \'Ujs\'aghy.
This work was supported by 
EC RTN2-2001-00440 ``Spintronics'', Projects OTKA D048665, T048782, T046303
and by the J\'anos Bolyai Scholarship.
\end{acknowledgments}


\begin{thebibliography}{99}

\bibitem{kondo}J. Kondo, Prog. Theor. Phys. {\bf 32}, 37 (1964).
\bibitem{hewson} For a review see A.C Hewson, 
{\em The Kondo Problem to Heavy Fermions}, Cambridge University Press (1993).
\bibitem{qd_th}
L.I. Glazman and M.E. Raikh, JETP Lett. {\bf 47}, 452 (1988);
T.K. Ng and P.A. Lee, Phys. Rev. Lett. {\bf 61}, 1768 (1988).
\bibitem{qd_ex} 
D. Goldhaber-Gordon, H. Shtrikman, D. Mahalu, D. Abusch-Magder, U. Meirav, 
and M.A. Kastner, Nature (London) {\bf 391}, 156 (1998);
S.M. Cronenwett, T.H. Oosterkamp, and L.P. Kouwenhoven, Science {\bf 281}, 
540 (1998);
J. Schmid, J. Weis, K. Eberl, and K. von Klitzing, 
Physica (Amsterdam) {\bf 256B-258B}, 182 (1998);
W.G. van der Wiel, S. De Franceschi, T. Fujisawa, J.M. Elzerman, 
S. Tarucha, and L.P. Kouwenhoven, Science {\bf 289}, 2105 (2000).
\bibitem{STM} J. Li, W.-D. Schneider, R. Berndt, and B. Delley, Phys. Rev. 
Lett. {\bf 80}, 2893 (1998);
V. Madhavan, W. Chen, T. Jamneala, M. F. Crommie, and N. S. Wingreen, 
Science {\bf 280}, 567 (1998);
 H. C. Manoharan, C. P. Lutz, and D. M. Eigler, 
Nature (London) {\bf 403}, 512 (2000). 
\bibitem{Balseiro}P. S. Cornaglia and C. A. Balseiro,
 Phys. Rev. B {\bf 66}, 115303 (2002); P. S. Cornaglia and C. A. Balseiro,
Phys. Rev. B {\bf 66}, 174404 (2002).
\bibitem{box}W. B. Thimm, J. Kroha, and J. v. Delft, 
Phys. Rev. Lett. {\bf 82}, 2143 (1999).
\bibitem{knight}
E. S. Sorensen and I. Affleck,
Phys. Rev. B {\bf 53}, 9153 (1996);
J. P. Boyce and C. P. Slichter, 
Phys. Rev. Lett. {\bf 32}, 61 (1974); Phys. Rev. B {\bf 13}, 379 (1976);
K. Chen, C. Jayaprakash, and H. R. Krishna-Murthy, Phys. Rev. Lett. {\bf 58},
929 (1987). 
\bibitem{persistent}
I. Affleck and P. Simon, Phys. Rev. Lett. {\bf 86}, 2854 (2001);
{\em ibid} {\bf 88}, 139701 (2002); 
H. P. Eckle, H. Johannesson, and C. A. Stafford, 
Phys. Rev. Lett. {\bf 87}, 016602 (2001); 
{\em ibid} {\bf 88}, 139702 (2002); 
P. Simon and I. Affleck, Phys. Rev. B {\bf 64}, 085308 (2001);
E. S. Sorensen and I. Affleck, Phys. Rev. Lett. {\bf 94}, 086601 (2005).
\bibitem{conductance}
P. Simon and I. Affleck, Phys. Rev. Lett. {\bf 89}, 206602 (2002).
\bibitem{Hand}T. Hand, J. Kroha, and H. Monien,
Phys. Rev. Lett. {\bf 97}, 136604 (2006). 
\bibitem{Wilson}
K.G. Wilson, \RMP{47}, 773 (1975);
H.R. Krishna-murthy, J. W. Wilkins, and K. G. Wilson \PRB{21}, 1003 (1980); \PRB{21}, 1044 (1980).
\bibitem{pert}
R. H. Bressmann and M. Bailyn, Phys. Rev. {\bf 154}, 471 (1967);
M. S. Fullenbaum and D. S. Falk, Phys. Rev. {\bf 157}, 452 (1967);
H. Keiter, Z. Phys. {\bf 223}, 289 (1969);J. Gan, 
J. Phys. Condens. Matter {\bf 6}, 4547 (1994)
\bibitem{barzykin}V. Barzykin and I. Affleck, Phys. Rev. B {\bf 57}, 432 (1998).
\bibitem{hirsch}J. E. Gubernatis, J. E. Hirsch, and D. J. Scalapino,
Phys. Rev. B {\bf 35}, 8478 (1987).
\bibitem{theo}T. A. Costi, A. C. Hewson, and V. Zlatic, J. Phys.:
  Condens. Matter {\bf 6}, 2519 (1994).
\bibitem{2imp} B. A. Jones, C. M. Varma, and J. W. Wilkins, Phys. Rev. Lett. 
{\bf 61}, 125 (1988); J. B. Silva, W. L. C. Lima, W. C. Oliveira, J. L. N. Mello, L. N. Oliveira, and J. W. Wilkins
Phys. Rev. Lett. {\bf 76}, 275 (1996)
\bibitem{Ralf}R. Bulla, Th. Pruschke, A.C. Hewson,
J. Phys.: Condens. Matter {\bf 9}, 10463 (1997); C. Gonzalez-Buxton and
K. Ingersent, Phys. Rev. B {\bf 57}, 14254 (1998); K. Chen and C. Jayaprakash,
Phys. Rev. B {\bf 52}, 14436 (1995).
\bibitem{ishi}H. Ishii, J. Low Temp. Phys. {\bf 32}, 457 (1978).
\end{thebibliography}
\end{document}